\documentclass{article}

\usepackage{amssymb}

\usepackage{graphicx}
\usepackage{epsfig}

\begin{document}

\title{Off line Parallax Correction for Neutral Particle Gas Detectors.}

\author{P. Van Esch}

\maketitle

\begin{abstract}
In a neutral particle gas detector, the parallax error resulting
from the perpendicular projection on the detection plane or wire
of the radial particle trajectories emanating from a point like
source (such as a scattering sample) can significantly spoil the
apparent angular resolution of the detector.  However, as we will
show, the information is not lost.  We propose an off line data
treatment  to restore as much as possible the original scattering
information in the case of a one-dimensional parallax effect.  The
reversibility of parallax follows from the algebraic structure of
this effect, which is different from the resolution loss which is
essentially irreversible.  The interplay between finite resolution
and parallax complicates the issue, but this can be resolved.
\end{abstract}

\section{Introduction}

\begin{figure}
  \includegraphics[width=8cm]{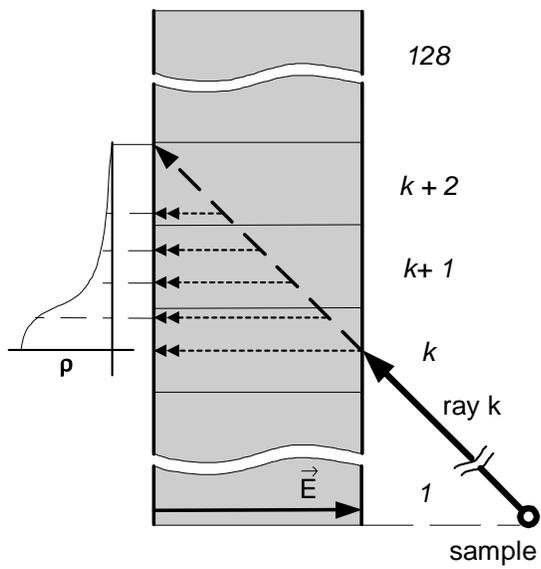}\\
  \caption{The principle of the parallax effect: a ray from the
    sample, hitting bin \(k\), also gives rise to a conversion
    probability in the next few bins.}\label{fig:principle}
\end{figure}

The parallax effect occurs when a point like source emits
radiation along spherically radial lines, which is then detected
by a detector with finite thickness and whose detection mechanism
projects this radial trajectory onto a plane or onto a line
(wire). This typically occurs in gas detectors of neutral
particles, as shown in figure \ref{fig:principle}: the neutral
particle follows its radial trajectory throughout the 'conversion
volume' where it suffers an exponentially distributed probability
to give rise to a cloud of electrical charge (to convert).  This
charge then follows electrical drift field lines until it reaches
a position sensitive detection element (such as a wire or a micro
strip detection element). Usually, for construction reasons, these
drift field lines are perpendicular to the detection surface or
line. So the conversion point (which is randomly distributed along
the radial line) will be projected perpendicularly onto the
detection element.  A narrow particle beam will thus give rise to
a smeared-out spot. This smearing-out effect is called the
parallax effect.  One tends to confuse it often with resolution
loss, but we will show in this paper that the algebraic structure
of the operator corresponding to 'parallax effect' is reversible,
while 'resolution loss' is essentially irreversible. We will also
work out the reverse operation in theory and in a simulated
example.

We haven't found much literature concerning this issue ; the idea
of inverting the parallax operator was suggested in
\cite{articleGrachev1996} in a JINR internal report; but it was
specified that the method gave problems when the detector had
finite resolution. Most people try to correct the detector
hardware to avoid parallax, such as \cite{articlechernenko1995a}
and \cite{articlechernenko1995b} by adapting the entire detection
volume, or \cite{brevetcomparat1988}, taken over by
\cite{articlebrookhaven1997}, by trying to correct for the
electric field projection.  An overview of hardware correction
techniques is given in the review \cite{articlecharpak1982}.

In this paper we will focus on the algebraic structure of the
parallax effect on the image, and how to inverse it, even in the
case of finite resolution. We will work out explicitly the case of
two geometries: a rectangular volume with projection perpendicular
on a plane, and a cylindrical volume with projection onto a wire.
In both cases, we only consider the parallax in one dimension. We
will subdivide this relevant dimension in 128 position bins, so an
image consists of 128 intensities.  We will have the 'true image'
\(I_{\textrm{true}}(n)\) which represents the intensity of
entering neutral particles in each bin (at the entrance window
level).  We have the conversion image \(I_{\textrm{para}}(n)\),
which is the perpendicular projection of the conversion points,
due to the parallax effect.  Finally, we have the raw image,
\(I_{\textrm{raw}}(n)\), which is the conversion image with the
finite resolution of the detection element.

\section{Algebraic structure of the parallax and resolution operators.}

\subsection{Resolution and convolution.}

In the continuum limit (an infinite number of infinitesimal bins),
finite resolution comes down to the convolution of the original
image with the point response of the detector describing its
resolution.  This point response is usually very well modelled by
a gaussian distribution but this can be any function with a single
extremum. This operation can be seen as the application of a
linear operator on the image function:
\begin{equation}
\label{eq:continuousconvolution} i_{\textrm{instrument}}(x) = \int
dy g(x-y) i_{\textrm{true}}(y)
\end{equation}
As it is well known, the translation invariance of this operator
results in eigenfunctions which are complex exponentials; in other
words, the operator is made diagonal by a Fourier transform. The
discrete version of this operation in the case of an infinite
train of samples, or a finite number of cyclic samples, is at the
heart of digital signal processing, as explained in
\cite{bookdsp}. We work however with a finite number of bins in
our detector, which are not to be considered cyclic.  As such,
Fourier representation (or the Z-transform) cannot really be
applied so we have to consider the operator on the image as a
general linear operator:
\begin{equation}
\label{eq:resolution} i_{\textrm{instrument}}(n) = \sum_{k=1}^N
g_{n,k} i_{\textrm{true}}(k)
\end{equation}
The square matrix \(g_{n,k}\) then consists of columns which are
given by:
\begin{equation}
\label{eq:resolutionmatrix}
 g_{n,k} = G(n-k)
\end{equation}
and \(G(l)\) is the sampled point response of the detector. There
is still of course a spectral representation: the eigenvalues of
the matrix \(g_{n,k}\).  This spectral representation is what
comes closest to the Fourier representation. As an example, let us
consider 128 bins with \(G(k) = e^{-\frac{k^2}{4}}\). This is a
gaussian point response that has a full width half maximum of 3.3
bins. The matrix \(g_{n,k}\) is a symmetrical matrix if the point
resolution function is symmetrical. Constructing the corresponding
matrix, and solving for the eigenvalues, we find the distribution
given in figure \ref{fig:eigenvalues}.  If this were a true
Fourier representation, we would talk about a low pass filter. The
ratio of the highest to the lowest eigenvalue is about \(10^4\),
which indicates that this is a highly irreversible operator in
practice. This is somehow normal: a low pass filter has
irreversibly cut away the high frequency components. In our case,
it has irreversibly cut away the projection onto the eigenvector
images corresponding to the small eigenvalues.

\subsection{The parallax operator.}

The parallax effect, even in the continuum limit, can not be
written as a convolution of the image with a kind of 'parallax
kernel function', because the parallax effect is strongly position
dependent.  But of course it is still a linear operator, so we can
write its effect always as:
\begin{equation}
\label{eq:parallax}
 i_{\textrm{instrument}}(n) = \sum_{k=1}^N
p_{n,k} i_{\textrm{true}}(k)
\end{equation}

\begin{figure}
  \includegraphics[width=8cm]{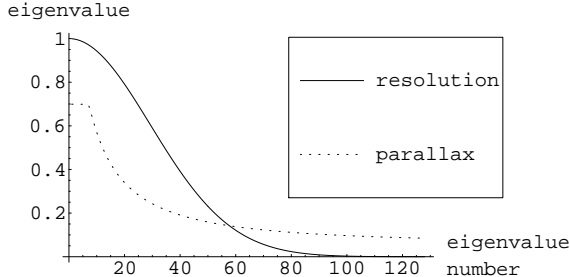}\\
  \caption{The eigenvalues of the operators \(G\) and \(P\) corresponding
    respectively to a gaussian point
  response \(G(k) \sim e^{-\frac{k^2}{4}}\) and the parallax in the case of
  a rectangular detection volume, over 128 bins, in descending order.}\label{fig:eigenvalues}
\end{figure}

Because even in the continuum limit this operator is not a
convolution, there is no remote resemblance to a Fourier
representation.   In order to look closer at the structure of the
matrix \(p\), we will have to think about the physical process
that is at the origin of the parallax effect. We will assume, as
shown in figure \ref{fig:principle}, that bin 1 of our 128 bins is
perfectly perpendicular to the sample ray, so that no parallax
occurs, and that the effect becomes stronger because the
projection direction deviates more and more from the radial
direction as the bin number increases. The general structure of
the matrix \(p_{n,k}\) is a sparse lower-triangular matrix, each
column containing the 'parallax tail' of a thin ray leaking in the
next few bins.  The diagonal elements are nothing else but the
fraction of particles in the ray hitting bin \(k\) that will also
convert in bin \(k\). This is determined by the geometry of the
conversion volume, its binning and the absorption law for the
neutral particles.  The diagonal elements, which are also the
eigenvalues for a triangular matrix, will not vary wildly. In
figure \ref{fig:eigenvalues}, the eigenvalues of such an operator
are displayed. \emph{It is from this observation that the
reversibility of the parallax effect results.}    Indeed, the
eigenvalues all having comparable values, the condition number of
the matrix will not be very high, and the inverse matrix will be
numerically very well defined.  A remark is maybe due: although
very regular, the matrix may be badly scaled, so some care is to
be taken in choosing the numerical inversion technique.  If the
detector had perfect resolution (meaning that \(G(m) \approx
\delta_{m,0}\)), then this would be the end of the story.

\subsection{Combination of resolution and parallax.}

In the previous two subsections we analyzed the algebraic
structure of the linear operator \(G\) that implements finite
resolution of the detector and the operator \(P\) that implements
the parallax effect. But in a real detector, both effects occur.
From the different structure of the eigenvectors, it is clear that
\(G\) and \(P\) do not commute.  We will assume here that the
finite resolution is an effect that occurs after the parallax
effect has had its effect, as  usually  is the case in neutral
(and even sometimes in charged) particle detectors: indeed, the
trajectory of the primary particle, before conversion into a
charge cloud, is usually only affected by the geometry (parallax).
Most other (resolution) effects (finite spread of charge,
diffusion, electronic noise etc...) occur afterwards.  We hence
have:
\begin{equation}
\label{eq:parallaxoperator} I_{\textrm{para}} = P
I_{\textrm{true}}
\end{equation}
and:
\begin{equation}
\label{eq:resolutionandparallaxoperator} I_{\textrm{raw}} = G
I_{\textrm{para}} = G P I_{\textrm{true}}
\end{equation}
Because \(G\) is nearly singular, we can't inverse the last
equation, so our hope to recover \(I_{\textrm{true}}\) from
\(I_{\textrm{raw}}\) vanishes.  But that is normal, we know that
we will suffer from the resolution loss, and the parallax effect
will certainly not improve on this.  However, we would like to
recover \(I_{\textrm{nopara}}= G I_{\textrm{true}}\), that is, the
image that has the resolution of the detector but doesn't suffer
from parallax anymore.  Formally, can write:
\begin{equation}
\label{eq:formalsolution}
 I_{\textrm{nopara}} = G (G P)^{-1}I_{\textrm{raw}}
\end{equation}
but again, because of the singular character of \(G\) this is not
workable as such.  It would have been if \(P\) and \(G\) commuted,
but they don't.  However, we can now apply a truncated singular
value decomposition on \(G P\) as explained in \cite{bookmatrix}:
\begin{equation}
\label{eq:svddecompositionG} G P \approx U^T \Sigma V
\end{equation}
where we limit the diagonal matrix \(\Sigma\) to those eigenvalues
of \(G P\) that are significant (it is our choice to specify what
we call significant: usually, given the numerical precision of the
data on which we will work, something of the order of 1\% or 10\%
will do).  What has been done here is replacing the numerically
near-singular matrix \(G P\) by a numerically regular matrix
\(\Sigma\) of lower rank (the number of accepted eigenvalues), and
two orthogonal projection operators \(U\) and \(V\).  As such, we
can work out equation \ref{eq:formalsolution}:
\begin{equation}
\label{eq:solution} I_{\textrm{nopara}} \approx I_{\textrm{sol}} =
G V^T \Sigma^{-1} U I_{\textrm{raw}}
\end{equation}
The solution proposed in equation \ref{eq:solution} is numerically
stable, but one should keep in mind that it is an approximation.
Indeed, eigenvectors of \(G P\) with a small eigenvalue (which are
thrown away in the truncated singular value decomposition) could
have large parts of eigenvectors of \(G\) with large eigenvalues,
hence contributing to the solution. However, there is no way to
calculate in a numerically stable way these contributions, so
there's more chance that they introduce spurious terms than
improve upon the solution. By using a truncated singular value
decomposition, we've thrown away all these potentially dangerous
contributions, but at the same time also their true contribution.

\section{Test cases.}

\subsection{Case of a rectangular drift volume.}

\begin{figure}
  \includegraphics[width=8cm]{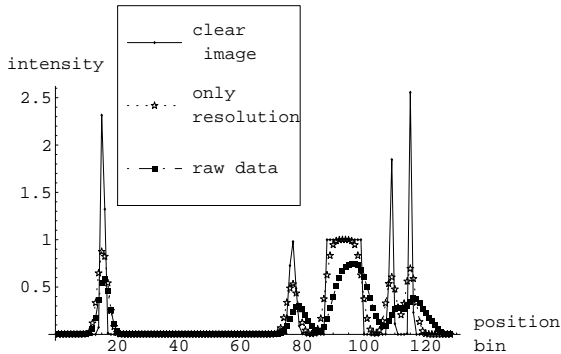}\\
  \caption{Test image and the response of the rectangular detector.
In the first case we only apply the finite resolution of the
detector, in the second case we apply first the parallax effect
and second the finite resolution, giving us the simulated raw
data.}
  \label{fig:rectangularresponse}
\end{figure}

\begin{figure}
  \includegraphics[width=8cm]{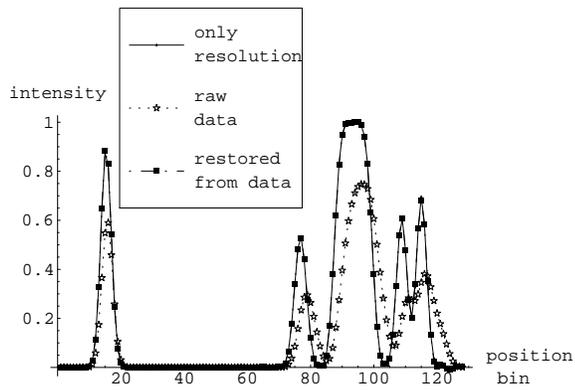}\\
  \caption{We compare the image the rectangular detector would have provided if
  it had finite resolution but no parallax effect, with the restored image
  using the raw data.}\label{fig:rectangularresult}
\end{figure}

To calculate the parallax operator in the case of a rectangular
drift volume, we consider a ray \(k\) that hits the entrance
window in the middle of the \(k^{\textrm{th}}\) bin and calculate
the distances \(s_j\) travelled on this ray from that point on to
each of the bin boundaries. The entry in the matrix for a
particular bin is then simply given by:
\begin{equation}
p_{k+j,k} = e^{-\mu s_{j}}-e^{-\mu s_{j+1}}
\end{equation}
This corresponds to the probability of conversion in bin \(k+j\).
We consider the example of a rectangular drift volume with depth
3cm, a height of 40cm divided into 128 position bins and a sample
distance of 40cm. The absorption coefficient is put to 0.4
absorption lengths per cm. The resolution of the detector (FWHM)
is 3.7 bins.  If we just generate a test picture, and see how the
picture is affected if (1) we only had the finite resolution of
the detector, or if (2) we had parallax and the finite resolution
of the detector, we obtain the images displayed in figure
\ref{fig:rectangularresponse}.  When we now apply our technique as
given in formula \ref{eq:solution}, even making an error and
assuming a better resolution of 2.9 bins instead of the simulated
3.7 bins, and using a relative cutoff of 5\% (so we reject
eigenvalues of the \(G P\) operator which are more than 20 times
smaller than the largest eigenvalue), we can restore the image.
Comparing it to the image we would have obtained with that
detector if it had finite resolution, but no parallax effect, we
obtain figure \ref{fig:rectangularresult}: the reconstruction is
almost perfect.

\begin{figure}
  \includegraphics[width=8cm]{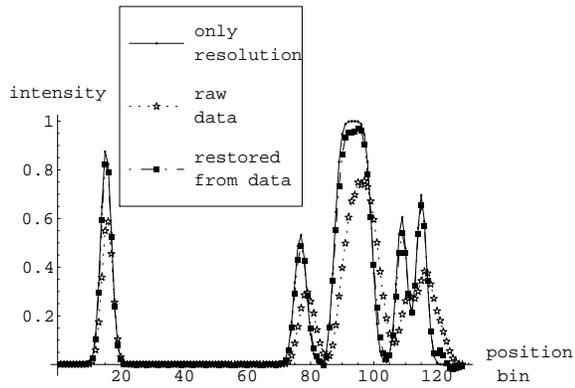}\\
  \caption{Reconstruction using slightly 'wrong' parallax and
  resolution parameters, and 2\% relative noise added to the raw
data.}\label{fig:rectangularresultnoise}
\end{figure}

In order to test the practical robustness of our technique, we
introduce errors on some parameters for the parallax operator used
during reconstruction.  The sample distance is increased from 40cm
to 43cm and the absorption coefficient is changed from 0.4 into
0.45 absorption lengths per cm. We also add random noise to the
raw data (we added a uniformly distributed noise of relative
intensity 2\%). We apply the technique keeping the tolerance to
5\%. We still obtain an image of reasonable quality, as shown in
figure \ref{fig:rectangularresultnoise}. Note that the small
overall decrease in amplitude is normal due to the different
absorption coefficients.  The peaks at the right hand side are
still well resolved, in the right position and of the correct
relative amplitudes and widths, which is usually the information
extracted in scattering experiments.

\subsection{Case of a cylindrical detector.}

We now consider a cylindrical geometry, with projection of the
charges perpendicular onto a central wire.  Although the geometry
is more complicated in this case, and although now we have to
consider not a ray (line) but a plane of radiation cutting through
the entire cylinder, the idea behind the calculation of the
parallax matrix elements is the same: a plane of radiation hitting
(at the front part of the cylinder) the middle of bin \(k\) will
give rise to a certain probability of detection in the next few
cells, and a geometrical calculation (together with an exponential
absorption law) will allow us to obtain the parallax operator.

We've applied this to the following case: a tube of radius 1.2cm,
of length 40cm, divided into 128 bins and a sample distance of
40cm. The absorption coefficient has been taken to be 0.4
absorption lengths per cm. We take the true resolution of the
detector to be (FWHM) 3.7 bins.

\begin{figure}
  \includegraphics[width=8cm]{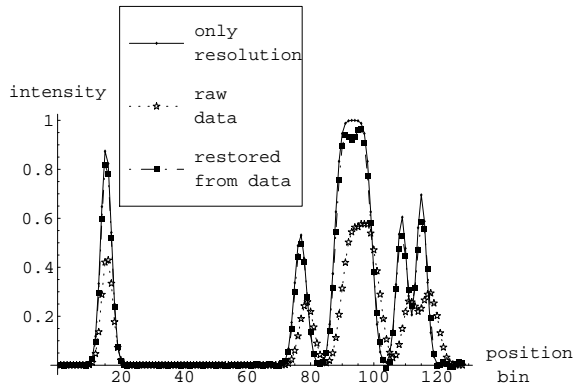}\\
  \caption{Cylindrical detector: the extracted picture from the raw data using our method is compared
  to what the detector would have done if there was only a finite resolution and
  no parallax effect.  For the reconstruction, we've introduced geometrical errors, absorption length
  errors and we added noise to the raw data.}
  \label{fig:cylindricalresult}
\end{figure}

Again introducing geometrical and detector parameter errors in the
reconstruction to test practical robustness,  we use an erroneous
resolution of 2.9 bins instead of 3.7, together with a sample
distance position of 42cm instead of 40cm, an absorption
coefficient of 0.45 absorption lengths per cm instead of 0.4, and
we add 2\% of relative noise to the raw data.  We use a cutoff of
the singular values at 5\% again. We then arrive at a restored
image as shown in figure \ref{fig:cylindricalresult}, which still
gives a satisfying result, and the same comments apply as in the
case of a rectangular detector volume.

\section{Discussion.}

After a theoretical explanation, we've tried to show, at the hand
of realistic case studies, that even a severe parallax effect does
not completely spoil the resolution of the image, and that useful
information can be extracted, using the proposed inversion
technique.  Sometimes some tuning of the tolerance allowed in the
truncated singular value decomposition is needed (especially in
difficult cases of strong parallax, badly known detector
parameters and noisy data) in order to get most out of the data.
We have to warn that we noticed that in the case of severe errors
on the detector parameters, or very high levels of statistical
noise, some small oscillations appear in the restored picture. So
this technique works best when we know rather well the detector
parameters and have data of relatively high statistical quality,
even if the parallax effect is strong. Of course preventing
parallax (using clever detector and instrument construction) is
always better than to cure it, but we tried to show in this paper
that a cure is possible for existing (or future) instruments and
data suffering from parallax.

\end{document}